\documentclass[preprint,showpacs,preprintnumbers,amsmath,english,amssymb]{revtex4}
\usepackage{graphicx}
\usepackage{dcolumn}
\usepackage{bm}

\begin{document}
\title{Current-driven vortex domain wall dynamics}
\author{J. He, Z. Li and S. Zhang}
\affiliation{Department of Physics and Astronomy, University of
Missouri-Columbia, Columbia, MO 65211}
\date{\today}

\begin{abstract}
Current-driven vortex wall dynamics is studied by means of a 2-d
analytical model and micromagnetic simulation. By
constructing a trial function for the vortex wall in the magnetic
wire, we analytically solve for domain wall velocity and
deformation in the presence of the current-induced spin torque. A
critical current for the domain wall transformation from the vortex
wall to the transverse wall is calculated. A comparison between the
field- and current-driven wall dynamics is carried out.
Micromagnetic simulations are performed to verify our analytical
results.
\end{abstract}
\maketitle

\section{Introduction}
Magnetic domain walls in magnetic films have various structural
forms which are determined by geometrical and material parameters.
For a magnetic wire with an ``infinite'' length and a finite width,
there are commonly two types of domain walls: transverse wall (TW)
and vortex wall (VW). Both of them are stable. Depending on the wire
thickness and width, one of the walls is usually more stable
\cite{McMichael,Miltat1}. However, in a certain range of the
parameters, the static energies of these two walls are comparable
and thus one can produce both types of walls in the same wire . By
using different initialization methods, one can create either wall
\cite{McMichael}. When a magnetic field or an electric current is
applied, both TW and VW are able to move along the wire. The
dynamics of the walls is generally very complex and micromagnetic
simulations are required in order to describe the details of the
domain wall motion. For the TW, a simplified and yet very insightful
analytical treatment was developed by Walker \cite{Walker}. A 1-d
wall profile, i.e., the magnetization direction in the wall depends
only on the coordinate along the wire, has been used to approximate
the TW profile. With this approach, one can analytically calculate
the wall velocity and wall distortion in the presence of magnetic
field and electrical current \cite{Li1}. For the VW, however, the
1-d wall profile fails to capture the wall structure and one needs
at least to use a 2-d model to approximately characterize the vortex
structure. In this paper, we propose such a 2-d model for the VW.
Our focus will be on the analytical calculation of the dynamic
behavior of the VW. Within our model, we are able to describe the
vortex wall motion, including the wall distortion, wall velocity and
wall structure transformation, in terms of material parameters and
external magnetic field or electric current. In particularly, we
show how the vortex core moves toward the edge of the wire when a
current or a field is applied along the wire. With a sufficiently
large current or field, the vortex core may vanish at the boundaries
of the wire edges and the transformation from the VW to the TW
occurs. This paper is organized as follows. In Sec.II, we develop
the analytical 2-d model for the VW. An equation of motion for the
domain wall is established. The steady state motions driven by the
current and by the field are investigated. We also compare the
dynamics between the TW and the VW. In Sec.III, the micromagnetic
simulations are performed. We compare the simulated results with the
analytical ones. Finally, we summarize the different features of
dynamics for TW and VW in Sec.IV.

\section{Analytical model}
\subsection{Equation of motion}
We start with the generalized Landau-Lifshitz-Gilbert equation
including the spin transfer torque terms \cite{Zhang}:
\begin{eqnarray}
\label{LLG} \frac{\partial {\bf M}}{\partial t}&=&-\gamma {\bf
M}\times \bf{H}_{\it eff}+\frac{\alpha}{\it{M}_{s}}{{\bf M}} \times
\frac{\partial {\bf M}}{\partial t}\\\nonumber
&&-\frac{b_{J}}{\it{M}^{\bf{2}}_{s}}{{\bf M}}\times\left({{\bf
M}}\times \frac{\partial{\bf M}}{\partial
x}\right)-\frac{c_{J}}{\it{M}_{s}}\bf{M}\times\frac{\partial
\textbf{M}}{\partial \it{x}}
\end{eqnarray}
where $\gamma$ is the gyromagnetic ratio, and ${\bf{H}}_{\it
eff}=-{\frac{\delta{W}}{\delta{\bf M}}}$ is the effective magnetic
field, and the $W$ is the total energy density which could be
written as $W = \it{A}\nabla^2{\bf m} +
(\it{K}/{\it{M}_{s}^2})({{\bf M}\times \bf{e}_{\it x}})^2 - \bf
{H_e}\cdot{\bf M} - (1/2)\bf H_d\cdot{\bf M}$, where $\it{A}$ is
exchange constant, $\it{K}$ is anisotropy, $\bf H_e$ is external
field, and $\bf H_d$ is magnetostatic field. $\alpha$ is the Gilbert
damping parameter, and $b_{J}=Pj_{e}\mu_{B}/eM_{s}(1+\xi^{2})$ and
$c_{J}=\xi b_{J}$, where $\emph{P}$ is the spin polarization of the
current; $j_{e}$ is the current density along the length direction
of magnetic wire; $\mu_{B}$ is Bohr magneton, $M_{s}$ is saturation
magnetization and $\xi$ (small, $\sim\alpha$) is a dimensionless
constant which describes the degree of the nonadiabaticity between
the spin of the nonequilibrium conduction electrons and local
magnetization.

To describe the motion of an entire domain wall, it is useful to
introduce a total force acting on the wall. Following Thiele \cite
{Thiele}, we define the total force
\begin{equation}
\label{F}\textbf{F}\{\theta,\phi\} \equiv \int dV \nabla W =
\int \left[
{\frac{\delta{\it{W}}}{\delta{\theta}}}(\bf{\nabla}\theta)+
{\frac{\delta{\it{W}}}{\delta{\phi}}}({\nabla}\phi)
\right]
\textit{dV}
\end{equation}
where $\theta$, $\phi$ are the angular components of $\bf{M}$ in the
spherical coordinate. For the steady-state motion of a domain wall,
we may write
$\theta=\theta({\bf{r}}-\emph{\textbf{v}}t)$,$\phi=\phi({\bf
r}-\emph{\textbf{v}}t)$, where $\emph{\textbf{v}}$ is the steady
velocity, then we have
\begin{equation}
\dot{\theta} = -\emph{\textbf{v}}\cdot\bf{\nabla}\theta,
\hspace{0.1in} \dot{\phi} = -\emph{\textbf{v}}\cdot\bf{\nabla}\phi.
\end{equation}
The above steady-state condition immediately reduces the
temporal-spatial differential equation, Eq.~(1), to a differential
equation with spatial variables only. By writing Eq.~(1) in the
angular components and by placing them into Eq.~(2), we obtain the
equation of motion for the domain wall \cite{Miltat2}:
\begin{equation}
\label{F}\bf{F}+\bf{G}\times(\emph{\textbf{v}}+\it{b}_{J}\hat{\bf{x}})+
\bf{\cal D} \cdot
(\alpha{\emph{\textbf{v}}}+\it{c}_{J}\hat{\bf{x}})=0
\end{equation}
where $\bf{G}$ is the domain gyrocoupling vector
\begin{equation}
\label{G} \bf{G}
=-\textit{M}_{s}\gamma^{-1}\int\textit{dV}\sin\theta(\bf{\nabla}\theta\times\bf{\nabla}\phi)
\end{equation}
and $\bf{\cal D}$ is the domain dissipation dyadic (tensor)
\begin{equation}
\label{D}\bf{\cal D} =
-\textit{M}_{s}\gamma^{-1}\int\it{dV}(\bf{\bf{\nabla}}\theta\bf{\bf{\nabla}}\theta+\sin^{2}\theta\bf{\bf{\nabla}}\phi\bf{\bf{\nabla}}\phi)
\end{equation}
The domain wall force defined in Eq.~(2) can be simplified when the domain wall
undergoes uniform motion. Let us separate the force in terms of
the internal force $\bf{F}^{\it{in}}$ and the external
force $\bf{F}^{\it{ext}}$,
\begin{equation}
\label{F}\bf{F}=\bf{F}^{\it{in}}+\bf{F}^{\it{ext}}.
\end{equation}
$\bf{F}^{\it{in}}$ contains all the forces from the internal energy
including anisotropy energy, exchange energy, and the magnetostatic
energy. When one sums over the internal energy contribution, the
total internal force vanishes due to Newton's third law. Therefore,
one may simply consider the external energy contribution to the
force in Eq.~(4) and (2), i.e.,
\begin{equation}
\label{F}\bf{F} =
\bf{F}^{\it{ext}}=\int\textit{dV} \left[
(\bf{\nabla}\theta)\frac{\partial}{\partial\theta}+(\bf{\nabla}\phi)
\frac{\partial}{\partial\phi} \right]
(-\bf{H}\cdot\bf{M}).
\end{equation}
where we have assumed that the external energy is solely from the external
field, $W=-{\bf H}\cdot {\bf M}.$ We will show later that we must consider
other external forces on the vortex wall when the wall reaches the boundary of
the wire. Note that if the profile of domain structures (i.e.
${\bf{M}(x,y,z)}$ in the moving frame of the steady motion) is
determined, the gyrocoupling vector, dissipation dyadic and static force
can be calculated from Eqs.~(5), (6) and (8), and the steady velocity
will be then readily derived from Eq.~(4).

In this Section, we shall apply the equation of motion, Eq.~(4), to
study the domain wall velocity of the vortex wall, driven by an
external magnetic field and by spin transfer torques. Let us first
consider a simplified head to head transverse wall as shown in
Fig.~(1a). For the transverse wall, we assume the wall profile can
be modeled by the Walker's trial function \cite{Walker},
\begin{equation}
\phi(x) = 2{\rm tan}^{-1} \exp\left(\frac{x}{\it{\Delta}}\right),
\hspace{0.2in} \theta(x)=\frac{\pi}{2}.
\end{equation}
where $\phi(x)$ and $\theta(x)$ are the angles between the direction
of the magnetization and the wire length direction ($+x$-axis) and
wire plane normal ($+z$-axis) respectively. $\it{\Delta}$ is the
domain wall width. By placing the above wall profile into
Eq.~(5)-(8), we find that the gyrocoupling vector is zero
$\bf{G}={0}$ and the dissipation dyadic has only one non-zero
component ${\cal D}_{xx}=-2\it{M}_{s}/{\gamma}\it{\Delta}$. The
external force is ${\bf{F}} = 2\it{H}\it{M}_{s}\hat{\bf{x}}$. By
inserting them into Eq.~(4), we immediately find the velocity:
\begin{equation}
\textit{v}_{x}=\frac{{\gamma}\it{H}\it{\Delta}}{\alpha}-\frac{c_{J}}{\alpha}.
\end{equation}
This result had been obtained previously \cite{Walker,Li1,Miltat2}.

It is difficult to analytically model the profile of the vortex wall
depicted in Fig.~(1b) by a single elementary function as we did for
the transverse wall. From previous simulation results
\cite{McMichael,Miltat1}, the VW structure might be understood as
two symmetrical transverse walls diagonally crossing the wire and a
central vortex core connecting the two TWs. It is noted that the two
TWs have opposite polarities, namely, the magnetization of the
centers of these two TWs orient in opposite directions. The
transitional region between these two TWs is sometimes called as the
Bloch line \cite{Bubble}, which characterizes the wall-polarity
reversal in analogy to the Bloch wall. For the VW we consider here,
this transitional region contains a vortex core. The magnetization
of the inner vortex core has a significant out-of-plane component
and thus the inner radius of the vortex core must be very small
since the out-of-plane magnetization enhances demagnetization energy
\cite{Tsuneto}. Outside the inner vortex core the magnetization lies
in the plane and the outer radius of vortex core is limited by the
transverse wall width and the wire width. To characterize the entire
VW profile, we separate the wall into \emph{three} parts: two
transverse walls and a vortex core (schematically shown in Fig.~(1c)
and  (1e)), and for the model in Fig.~(1e), they will be assigned to
different trial functions given below . For the vortex part
\cite{Shima,Tsuneto,Mertens,Oleg},
\begin{equation}
\left\{\begin{array}{ll} \theta=\left\{\begin{array}{clcc}
2\tan^{-1}\left(\frac{\sqrt{x^2+y^2}}{r_{core}}\right),\hspace{0.1in}(0\leq{x^{2}+y^{2}}<r_{core}^2),\\
\\
\frac{\pi}{2},\hspace{1in}(r_{core}^2\leq{x^{2}+y^{2}}<R^2),
\end{array}\right.
\\
\\
\phi=q \cdot \it{arg}(x +
iy)+c\frac{\pi}{2},\hspace{0.5in}(0\leq{x^{2}+y^{2}}\leq{R^{2}}).
\end{array}\right.
\end{equation}
where $r_{core}$ and $R$ are the inner-core and outer radius of the
vortex respectively, $q(=\pm1,\pm2...)$ is the vorticity of the
vortex, $c(=\pm1)$ is the chirality of the vortex, and
$i=\sqrt{-1}$. Here we just use the $\it{arg}$ function of  complex
variable $x + iy$ as a convenient way to express $\phi$
\cite{Mertens}. In this paper, we only consider a single vortex
($q=1$). And we note that the vortex profile we introduced in
Eq.~(11) does not include its image profile \cite{Mertens}. For the
wire structure, the image profile would consists of a series of
terms formed by multiple reflections of two wire boundary planes.
Including these images would make the analytical calculation
intractable. Instead, we introduce a `restoring force' \cite{kappa}
due to the induced charges along the boundaries of the nanowire.
This restoring force has similar roles of the `image vortices' by
the method of image \cite{Mertens}. For the two transverse walls
\cite{Walker,Li1} where ($-\infty<x<+\infty$),
\begin{equation}
\left\{\begin{array}{ll} \theta={\pi}/{2},\hspace{1in}(-y_{0}\leq{y}\leq{y_{0}}),\\
\\
\phi=\left\{\begin{array}{clcc}
2\tan^{-1}\exp\left(\frac{x+x_0}{w''}\right),&(-y_{0}-\delta y\leq{y}\leq{0}),\\
\\
2\tan^{-1}\exp\left(\frac{x-x_0}{w'}\right),&(0\leq{y}\leq{y_{0}-\delta
y})
\end{array}\right.
\end{array}\right.
\end{equation}
where $x_{0}$ is the distance from the centers of each TW part to
the vortex core center, $w'$ and $w''$ are the wall widths of two
TWs, $y_{0}$ is the half width of the magnetic wire, $\delta y$ is
the displacement of vortex core away form the wire center in the
y-direction. We realize that the TW and vortex core defined above
are not strictly valid because the TW is extended to the vortex
region when $0\leq{x^{2}+y^{2}}\leq{R^{2}}$. However, this
ill-defined overlapped region does not contribute significant errors
since $\nabla\theta$ and $\nabla\phi$ is exponentially small in this
region. By using the above wall profile, we are able to calculate
$\bf{G}$, $\bf{\cal D}$ and $\bf{F}$ defined in Eqs.~(5), (6), and
(8) in a similar way as the calculations in \cite{Huber, Slon1}. The
steady-state velocity can then be calculated from Eq.~(4).

\subsubsection{Current-driven domain-wall motion}
We consider the current-driven vortex wall motion without the
external field so that ${\bf F}={0}$, see Eq.~(8). Let us apply the
equation of motion, Eq.~(4), to the vortex part of the wall and
neglect its interaction with two TW parts. By placing Eq.~(11) into
Eq.~(5) and (6), we find the explicit expressions for $\bf{G}$ and
$\bf{\cal D}$:
\begin{equation}
{\bf{G}}=-2{\pi}\it{M}_{s}\gamma^{-1}\it{p}\hat{\bf z}=G_{v}\hat{\bf
z}
\end{equation}
\begin{eqnarray}
{\bf{\cal D}}&=&-M_{s}\gamma^{-1}{\pi}\ln\frac{R}{a}(\hat{\bf
x}\hat{\bf x} +\hat{\bf y}\hat{\bf y})\\\nonumber &=&{\cal
D}_{v}(\hat{\bf x}\hat{\bf x}+\hat{\bf y}\hat{\bf y})
\end{eqnarray}
where $\hat{\bf x}$, $\hat{\bf y}$, $\hat{\bf z}$ are unit vectors
in the directions of the wire length, width and thickness,
respectively. The $p(=\pm1)$ represents the polarity of the vortex
core ($\theta(0)=0,\hspace{0.05in}or,\hspace{0.05in}\pi$), and $a$
is the lattice constant of the crystal structure of wire material.
Note that in the calculation of ${\bf{\cal D}}$, we have neglected a
small part of the vortex core ($\thicksim a^2$), and this does not
considerably effect the wall dynamics \cite{Huber}. By placing
Eqs.~(13)-(14) and ${\bf F}={0}$ into Eq.~(4), we have
\begin{equation}
\left\{\begin{array}{rl}
-G_{v}\it{v}_{y}+{\cal D}_{v}(\alpha\it{v}_{x}+c_{J})=0\\\\
G_{v}(\it{v}_{x}+b_{J})+\alpha{{\cal D}_{v}}\it{v}_{y}=0
\end{array}\right.
\end{equation}
and one immediately finds:
\begin{equation}
\left\{\begin{array}{ll} \it{v}_{x_0}=-b_{J}
\frac{G_{v}^{2}+\alpha\xi{\cal D}_{v}^2}{G_{v}^{2}+\alpha^{2}{\cal D}_{v}^2}\cong-b_{J}\hspace{0.1in}(\alpha\ll1,\xi\ll1)\\\\
\it{v}_{y_0}\cong-\frac{1}{2}b_{J}p(\alpha-\xi)\ln{\frac{R}{a}}
\end{array}\right.
\end{equation}
It is shown that the velocity of VW has two components:
$\it{v}_{x_0}$ along the wire and $\it{v}_{y_0}$ perpendicular to
the wire. The nonzero $v_{y_0}$ (except $\xi=\alpha$) is caused by
the gyrotropic term (nonzero $G$) of the vortex dynamics
\cite{Thiele}. The sign of $\it{v}_{y_0}$ (to which edge of the
wire) is determined by the sign of the product $p(\alpha-\xi)$,
i.e., it is vortex polarity $p$ dependent. If we characterize the
vortex by a Bloch line, the perpendicular motion might be understood
in terms of the moving Bloch line inside and along the wall surface
\cite{Bubble}.

We recall that the velocity of the TW is  $v_{x}= -c_J/\alpha$ and
$v_y = 0$ in the steady-state motion. The different velocities for
the vortex core and the transverse wall can not be a steady state
solution of the entire wall. In fact, the interaction between these
different parts of the wall must be included in order to reach a
common velocity in the steady state. One can immediately see that
the motion of the vortex core along the $y$-direction would push one
TW and pull the other. As a result, one TW expands and the other
gets compressed, see Fig.~1(e). Conversely, these distorted TWs
would produce a reacting force to the vortex core; sometimes it is
called a restoring force \cite{Miltat2,Ivanov,Shima}. If we model
the forces between the vortex core and the TW by elastic potentials
in analogy to spring-connected particles, we must added the spring
force as an external force when one applies Eq.~(4) to each
individual parts of the vortex wall. Then, two scenarios are
possible: the reacting force is strong enough to completely halt the
perpendicular wall velocity; or the reacting force is unable to stop
the vortex from colliding to the wall edge. In the former case, the
perpendicular wall velocity eventually reaches zero and a
steady-state wall velocity along the wire is achieved. The final
wall velocity is precisely same as that of the transverse wall
(i.e., $v_x = -c_J/\alpha$ and $v_y = 0$, see calculations below).
In the latter case, the vortex core collides with the wire edge and
vortex core can either vanishes (move out of the wire) or be
reflected, i.e., the transformation of the vortex wall to other
types of the walls occurs. Usually, the vortex core disappears at
the edge and the wall becomes a TW \cite{Klaui}.

When a steady-state motion is achieved, we may apply Eq.~(4) to the
entire wall. The interaction between the vortex core and the TW
walls becomes internal force and it does not enter Eq.~(4).
However,there is an external force acting on the entire wall, which
is due to the magnetic charges coming from the distorted TW parts at
the boundaries. This force can be again analogy to the counteractive
force from the fixed boundaries to the compressed springs. If we
model this force linearly depending on the displacement of the
vortex core $\delta \it{y}$ in the $y$-direction, i.e.,
$\it{F^{re}}=-{\kappa \delta \it{y}}$ where the constant $\kappa$ is
discussed in Ref.~\cite{kappa} , we find, from Eq.~(4),
\begin{equation}
\left\{\begin{array}{rl}
-G_{v}\it{v}_{y}+{\cal D}_{xx}(\alpha\it{v}_{x}+c_{J})=0\\\\
-\kappa\delta\it{y}+G_{v}(\it{v}_{x}+b_{J})+\alpha{{\cal
D}_{yy}}\it{v}_{y}=0
\end{array}\right.
\end{equation}
where we have used and determined
\begin{equation}
{\bf{G}} = G_{v}\hat{\bf z}
\end{equation}
\begin{eqnarray}
{\bf{\cal D}}&=&({\cal D}_{v}+{\cal D}_{t})\hat{\bf x}\hat{\bf
x}+{\cal D}_{v}\hat{\bf y}\hat{\bf y}\\\nonumber &=&{\cal
D}_{xx}\hat{\bf x}\hat{\bf x}+{\cal D}_{yy}\hat{\bf y}\hat{\bf y}
\end{eqnarray}
where ${\cal D}_{t}=-2M_{s}\gamma^{-1}(\frac{y_0-\delta
y}{w'}+\frac{y_0+\delta y}{w''})$. Note that the TW parts have no
contribution to the gyrocoupling vector $\bf{G}$, however they
modify the dissipation dyadic with ${\cal D}_{t}$.

In the steady state of motion, $v_y=0$, we immediately find from
Eq.~(17)
\begin{equation}
v_x = - \frac{c_J}{\alpha}
\end{equation}
and
\begin{equation}
\delta\it{y}= \frac{G_{v} b_J}{\kappa} \left( 1- \frac{\xi}{\alpha}
\right)
\end{equation}

The above simplified analysis shows that the steady state domain
wall velocity is a universal constant $v_x = -c_J/\alpha$,
independent of the wall structure, {\em as long as the restoring
force from the edges of the wire is strong enough to halt the wall
motion in the direction perpendicular to the wire}. We can
qualitatively estimate the critical current $b_J^c$ when the
restoring force just barely prevents the wall colliding with the
wire edge, i.e., when $\delta y=y_0$ ($y_0$ is the half width of the
wire). By using Eq.~(21), we have
\begin{equation}
b_J^c =  \frac{\kappa y_0}{G_v} \left( 1- \frac{\xi}{\alpha} \right)
^{-1}
\end{equation}

If the current is larger than the above current density, the
restoring force is unable to stop the vortex core colliding to the
boundary and the VW transforms into a TW. The wall transformation
may be qualitatively understood in terms of the conservation of the
topological charges \cite{Oleg}: when the vortex core reaches the
edge, the vortex with the winding number $+1$ is absorbed by one of
the edge defects (one of the transverse wall parts) with winding
number $-1/2$, then a $+1/2$ edge defect must be created to conserve
the topological charges. This $+1/2$ edge defect together with the
$-1/2$ edge defect on the other side of wire edge is equivalent to a
simple transverse wall with a uniform polarity.

\subsubsection{Field-driven domain-wall motion}
By adding the external force due to the magnetic field (defined in
Eq.~(8)) to Eq.~(17), we can similarly derive the domain-wall steady
motion velocity. In the steady state, the external force on the
entire wall is
\begin{equation}
{\bf{F}}=4\it{y}_{0}\it{H}\textit{M}_{s}\hat{\bf
x}+c\pi\textit{H}\textit{M}_{s}(R-a)\hat{\bf y}=F_{x}\hat{\bf
x}+F_{y}\hat{\bf y}.
\end{equation}
In the absence of the current where $c_J=b_J=0$, we find that from
Eq.~(17) the steady-state velocity along the wire ($v_y =0$) is
\begin{equation}
\it{v}_{x}=-\frac{F_{x}}{\alpha{{\cal
D}_{xx}}}=\frac{\gamma{H}y_{0}}{\alpha}\left[\frac{4}{\pi\ln\frac{R}{a}+2(\frac{y_0-\delta
y}{w'}+\frac{y_0+\delta y}{w''})}\right]
\end{equation}
and the displacement of the vortex center in the direction of the
wire width is
\begin{equation}
\delta y=\frac{G_v\it{v}_{x}+F_{y}}{\kappa}
\end{equation}

As in the case of TW, the wall velocity is not universal and it
depends on the details of the wall structure. Comparing with TW, the
velocity of the VW is smaller \cite{Slon2}; this can be
qualitatively understood in terms of a smaller wall width for the
VW. Let us take the width of the TW part \cite{Miltat1} as
$\frac{y_{0}-\delta y}{w'}=\frac{y_{0}+\delta y}{w''}=\pi$
 (again, $y_{0}$ is half width of the wire) . And the
width of the TW ($\Delta$ in Eq.~(9)) in the same wire is also taken
as $\Delta=2y_{0}/\pi$.  For the VW, we may define an effective wall
width from Eq.~(24)
\begin{equation}
w_{vw} = \Delta \left(\frac{1}{2+ \frac{1}{2} \ln
\frac{R}{a}}\right)
\end{equation}
For example, for a wire $R=w/2= 20 nm$ and the lattice
constant $a=0.5 nm$, the ratio of the TW to VW velocity is about $2+
\frac{1}{2} \ln \frac{R}{a} = 3.8$.

When the magnetic field is large, the restoring force is unable to
stop the wall colliding with the wall edge. The vortex wall will be
annihilated and the transformation to the TW occurs. Similarly, we
may estimate the critical field $H_{e}^c$ for the wall
transformation by the condition $\delta y=y_0$. We thus find
\begin{equation}
H^c = \kappa y_{0}
\left[\frac{{\gamma}G_{v}w_{vw}}{\alpha}+c{\pi}M_s(R-a)\right]^{-1}
\end{equation}

\section{Micromagnetic Simulation}
\subsection{Simulation procedure}
To verify the analytical results derived above, we perform
micromagnetic simulations by directly solving Eq.~(1) for a
defect-free magnetic wire sample. The geometrical size of the wire
is $2{\mu}m$ long (x direction), 128nm wide (y direction) and 8nm
thick (z direction). The grid size is taken as
$4\times4\times8nm^3$. The material parameters are: the exchange
constant $A=1.3\times10^{-6} erg/cm$, the anisotropy field $H_{K} =
0 (Oe)$, the saturation magnetization $M_s= 800 emu/cc$, the spin
polarization $P=0.5$, the damping parameter $\alpha$ and the
non-adiabaticity factor $\xi$ varying from $0.01$ to $0.08$.

The magnetization at both ends is set to be along the x-direction
and direct inward to the wire and we use free boundary conditions
for other dimensions. Since the domain wall may move several
micrometers in some cases, it is important to keep the domain wall
far away from the ends to reduce the influence of magnetostatic
fields from the wire ends: we shift the $x$-coordinate of the entire
domain wall after each numerical iteration so that the center of the
wall is always located at the center of the wire. By using two
initial domain configurations similar to \cite{McMichael}, we
generate two types of walls in the middle of the wire, the TW in
Fig.~(1a) and the VW. The VW has four equivalent configurations
corresponding to different polarity ($\it{p}=\pm1$) and chirality
($c=\pm1$). We only show the VW for $\it{p}=-1$ and $c=1$ in
Fig.~(1b). The TW and VW shown in Fig.~(1) are our initial wall
configurations. At time $t=0$, a constant spin torque or a magnetic
field is applied to the wire, and we calculate the wall motion as a
function of time until the wall reaches steady motion or until a
sufficiently long time that the wall motion remains non-uniform.

\subsection{Simulation results}


To compare the analytical results on the universal domain wall
velocity, we first consider applying a sufficiently small spin
torque ($b_J=25(m/s)$) so that the vortex core remains inside the
wire. We vary damping parameter $\alpha$ and non-adiabaticity
coefficient $\xi$ from 0.01-0.08 to determine the relation between
the domain wall velocity and these two parameters. In Fig.~(2), We
have found that the initial velocity of the VW in the $x$-direction
is always $b_J$ (independent $\alpha$ and $\xi$), while the total
displacement of the center of the wall in the $y$-direction
quantitatively agrees with Eq.~(21). As expected, the $y$-component
velocity gradually reduces to zero when the center of the core moves
toward the edge. The restoring force eventually stops the wall
motion in the $y$-direction, and a steady state motion with the
velocity $v_x = -c_J/\alpha$ along the $x$-direction is reached. In
Fig.~(3), we show the relations of the terminal velocity $v_x$ with
both $\alpha$ and $\xi$. These simulated results are in excellent
agreement with our analytical formula.


When we increase the applied current density (or the strength of the
spin torques) above a critical value, the simulations show the
vortex core moves outside of the nanowire from the edge and the VW
transforms to TW. The transformed TW moves at the velocity
$-c_J/\alpha$ \cite{He}. We note that the critical current density
depends on the parameters. A special case is when $\xi=\alpha$. We
find that the VW can maintain its wall structure up to a much higher
current density. This is because the vortex wall does not move
toward the wire edge, see Eq.~(21). However, this accidental case
should not be considered as a general property of the VW motion.
When the current density is further increased, the TW will also
deform during its motion and the velocity is highly non-uniform,
known as Walker's breakdown \cite{Walker}. The alternative
appearance of transverse wall and vortex wall (or anti-vortex wall)
\cite{Miltat3} is seen in simulations (not shown here). In our
analytical calculations, we exclude these extremely complicated wall
dynamics.


As a comparison, we also simulate the field-driven domain wall
motion at sufficient small fields. The behavior of the VW driven by
field is similar to the current-driven case: namely, besides the
motion along the wire, the vortex core moves toward the edge and
eventually stops due to the restoring force. At a critical field,
the VW overcomes the restoring force and the VW transforms into TW.
While the steady velocity of TW and VW are both proportional to the
magnetic field $H$ and $\alpha^{-1}$, see Fig.~(4), the VW has a
smaller velocity compared to the TW for the same field.

Finally, we point out that the analytical and simulation analysis in
this paper is for a defect-free wire. In a realistic wire, the
defects are unavoidable. We expect that our analytical results may
not be applicable. For example, the steady state wall velocity may
not be universal and the defect pinning depends on the detail of the
wall structure. Indeed, we have seen in our earlier simulation
\cite{He} that the critical current is larger for depinning a TW
than for a VW. An even more remarkable finding is that the VW
firstly begins depinning, then transforms to a TW, which is finally
pinned by defects again (wall stops). The phenomenon occurs when one
applies a moderate current density between the critical values of
two depinning currents of the VW and TW \cite{He}. The above feature
has been captured in recent experiment \cite{Klaui}.

\section{Conclusion}
In summary, we have calculated the velocity of the VW in the
magnetic wire. In a defect-free wire, the terminal velocity is
independent of the wall structure for the current-driven steady
motion, while for the field-driven case, the effective domain-wall
width is a key parameter for the domain-wall mobility. The
transformation between the VW and the TW is explained as the
consequence of the perpendicular motion of the vortex core and the
conservation of topological charges. Our model is further supported
by numerical solutions and in agreement with experimental results.
The research was supported by NSF (DMR-0314456 and ECS 0223568).

\pagebreak
\bigskip
\noindent {Figure Caption}

\bigskip

\noindent{FIG.1} Magnetization patterns of (a) a transverse wall and
(b) a vortex wall, without an applied current or field. (c)
Schematic model for the vortex wall (b): a vortex wall is modeled by
two TWs and a vortex core, shown in gray areas. The vortex core is
at the center of the wire and $\pm{x_0}$ are the positions of
centers of the TWs. $R$ is the outer radius of the vortex core, $w$
is the wall width of two TW walls, and $y_0$ is the half width of
the wire. (d) The magnetization pattern of the vortex wall in the
presence of the current, which is calculated by micromagnetic
simulation in the Sec.III. The parameters are: the spin torque
$b_J=25(m/s)$, and damping parameter $\alpha=0.05$, non-adiabaticity
coefficient $\xi=0.04$. (e) The schematic model for the vortex wall
(d). $\delta y$ is the displacement of the vortex toward the edge of
the wire. The wall structure is not symmetrical, and one TW is
expanded and the other is compressed.

\bigskip

\noindent{FIG.2} The dependence of (a) x-component of initial
velocity $v_{x_0}$ and (b) the y-component displacement $\delta y$
as a function of the nonadiabaticity parameter $\xi$ for two
different polarity VWs ($p=\pm1$). We have used $b_J=25(m/s)$ and
$\alpha=0.05$. The straight lines are simple data fits.

\bigskip

\noindent{FIG.3} Steady-state velocity of TW and VW as a function of
(a) the damping parameter, and (b) the non-adiabaticity parameter.
The current density is $b_J=25 m/s$. The straight lines are simple
data fits.

\bigskip

\noindent{FIG.4} Steady-state velocities of both \emph{TW }and VW as
a function of (a) the magnetic field and (b) the damping parameter.
The current density is $b_J=25 m/s$. The straight lines are simple
data fits.

\newpage

\begin{figure}
\centering
\includegraphics[width=10cm]{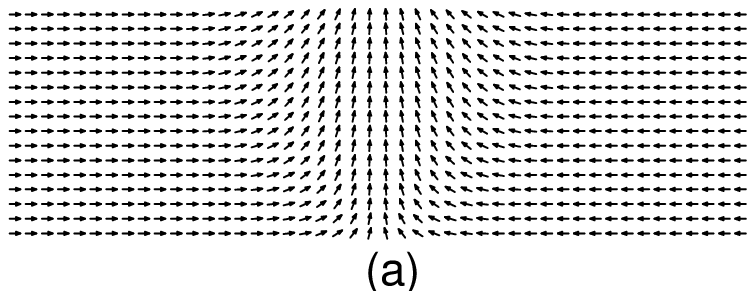}
\includegraphics[width=10cm]{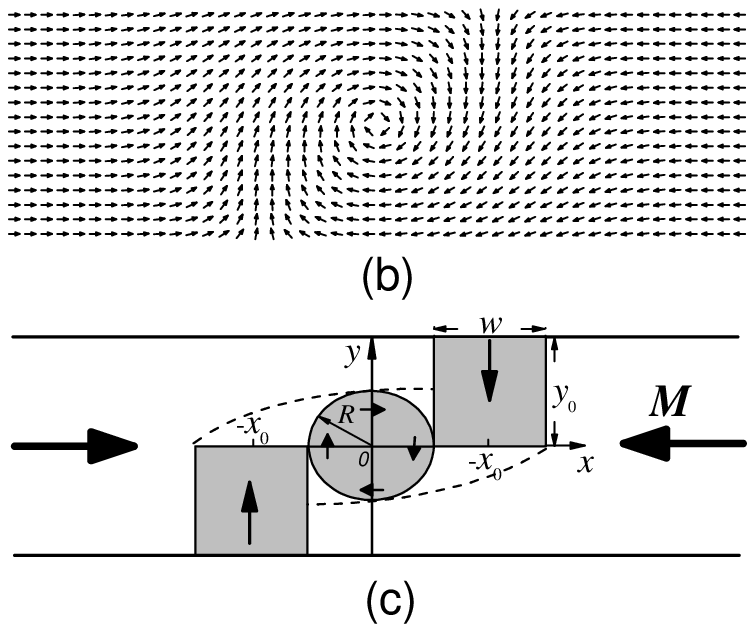}
\includegraphics[width=10cm]{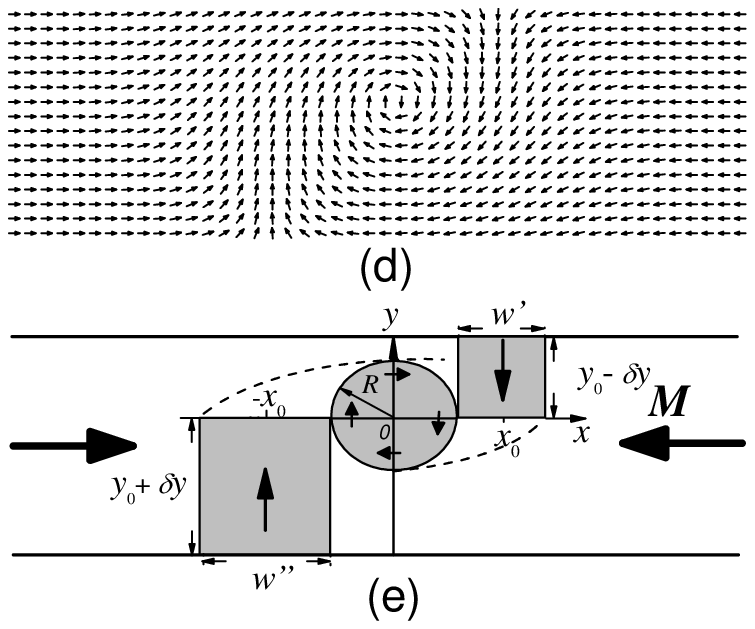}
\caption{}
\end{figure}

\bigskip
\pagebreak
\newpage

\begin{figure}
\centering
\includegraphics[width=14cm]{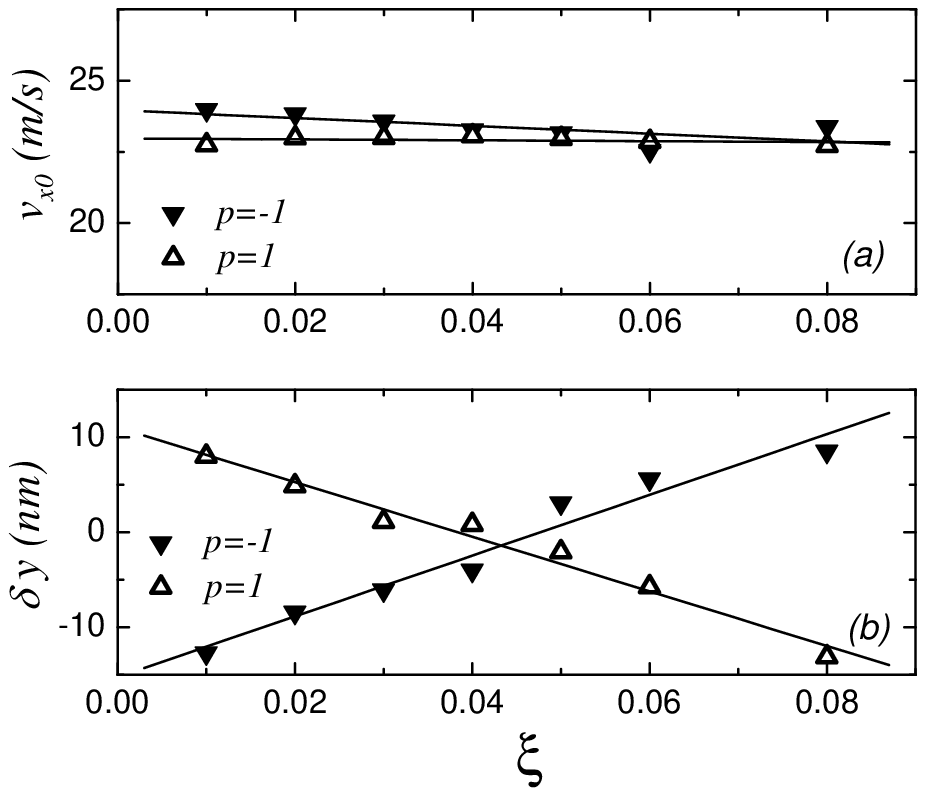}
\caption{}
\end{figure}

\bigskip
\pagebreak
\newpage

\begin{figure}
\centering
\includegraphics[width=14cm]{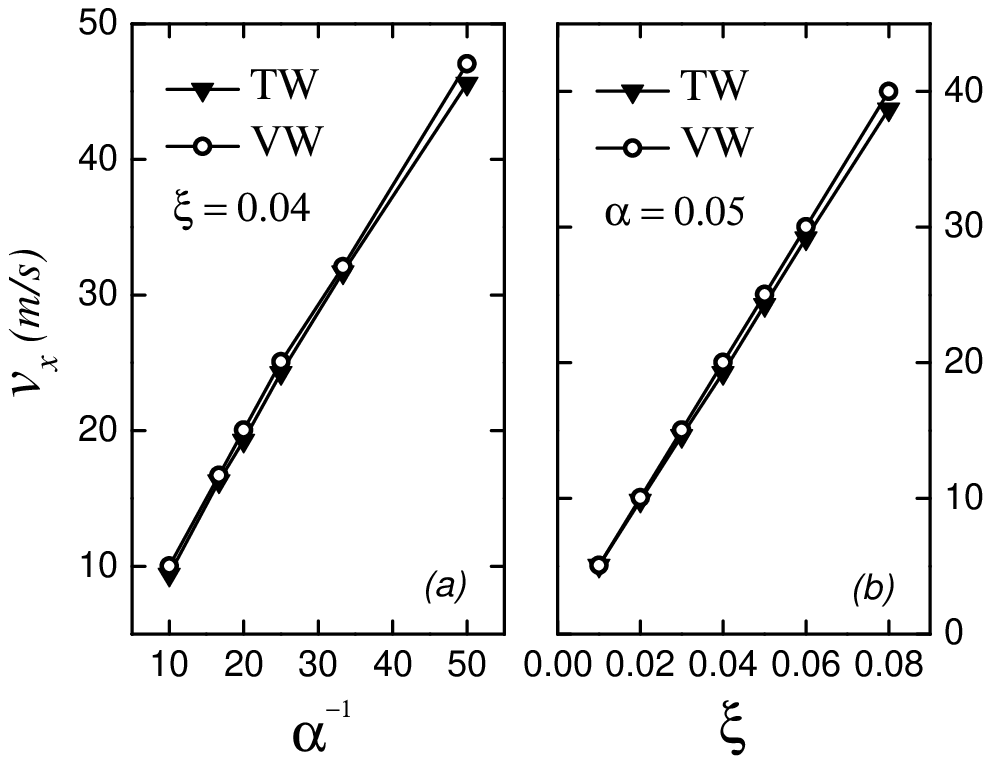}
\caption{}
\end{figure}
\bigskip
\pagebreak
\newpage

\begin{figure}
\centering
\includegraphics[width=14cm]{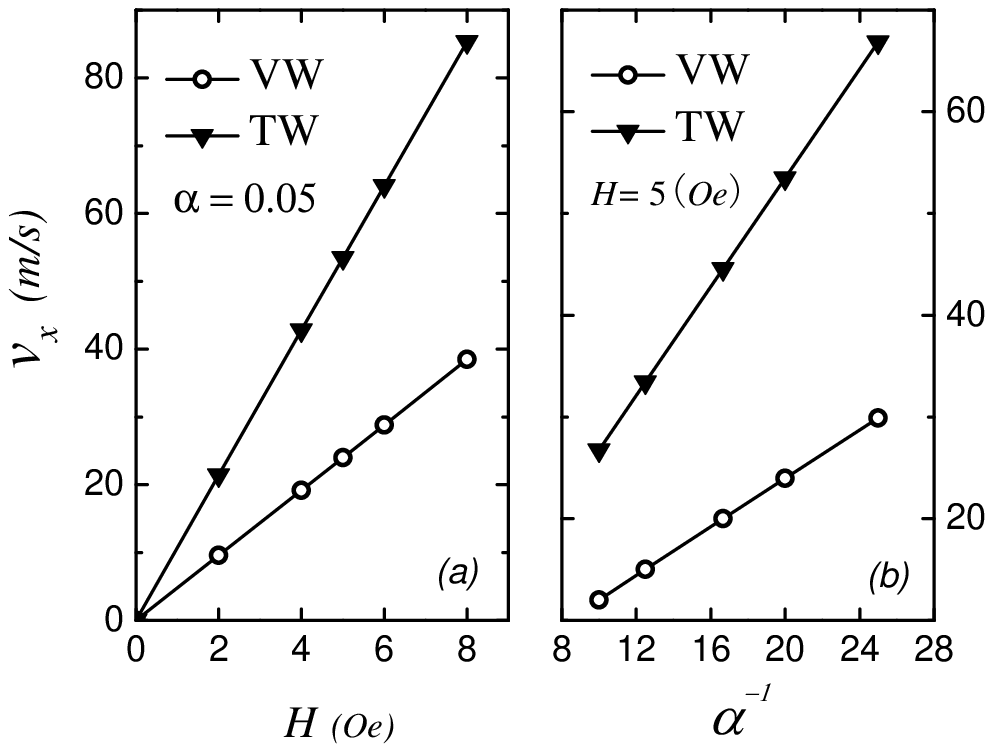}
\caption{}
\end{figure}

\begin{thebibliography}{1}
\bibitem{kappa}
While the calculation of the restoring force depends on details of
the wall structure, one can estimate it in the following. Consider a
simple TW in the wire, the wall energy $E(w) = \frac{2A}{w}+2Kw$,
where $A$ is the exchange constant, $w$ is the wall width and $K$ is
anisotropy constant. At equilibrium, the wall width $w_0$ is
determined by $\frac{\partial E}{\partial w_0}=0$, i.e., $w_0 =
\sqrt{A/K}$. When the wall width is compressed or expanded,  the
wall energy can be expanded, up to the second order of $w-w_0$,
$E(w) = E(w_0) + \frac{2K}{w_0}(w-w_0)^2$. Equivalently, the
restoring force due to the wall width change is
$F=-\frac{4K}{w_0}(w-w_0)$. In the present geometry, we approximate
the width as $\frac{y_{0}-\delta y}{w'}=\frac{y_{0}+\delta
y}{w''}=\pi$ \cite{Miltat1}. We take $\delta y = \pi \delta w$ , the
restoring force due to the displacement of the vortex core in
$y$-direction is then $F^{re}=-\kappa \delta y$, where
$\kappa=\frac{4K}{\pi w_0}$. When the vortex reaches the boundary,
the magnetostatic energy contribution due to the surface charges
along the boundary must be included in evaluating $\kappa$. A
detailed calculation can be found in Ref.~\cite{Shima}.
\end{thebibliography}
\end{document}